\documentclass[twocolumn,twoside]{article}
   \usepackage{graphics}
   \newcommand{\kms}{$\mathrm{\,km\,s}^{-1}$}
   
   \newcommand{\Vt}{$\mathrm{V_{t}}$}
   \newcommand{\Teff}{${T_{\rm eff}}$}
   \newcommand{\logg}{$\log \mathrm{g}$}

   \newcommand{\h}{\linebreak \hspace*{3ex}}
   \newcommand{\hb}{\\ \hspace*{2ex}}
   
\begin{document}
   \title{DISCOVERY OF THE FIRST SUPER-LITHIUM RICH \\
           BEAT CEPHEID: V371 PER}
   \author{V.~V.~Kovtyukh$^1$, N.~I.~Gorlova$^2$, \&  M.~Hillen$^2$\\[2mm]
   \begin{tabular}{l}
   $^1$ Astronomical observatory of I.~I. Mechnikov Odessa National University\hb
   T.G.\,Shevchenko Park, Odessa 65014 Ukraine, {\em val@deneb1.odessa.ua}\\
   $^2$ Institute of Astronomy, Celestijnenlaan 200D, 3001, Leuven, Belgium\\
   \\[2mm]
   \end{tabular}
   }
   \date{}
   \maketitle
   \indent
   ABSTRACT.
  Four high-resolution spectra of the double-mode Cepheid V371 Per,
  obtained for the first time, showed the
  presence of the abnormally strong Li {\sc i} 6707.76 \AA\ line.
  Our analysis of the light element abundances
  indicates that the star did not go through the evolutionary dredge-up stage.
  Large distance from the galactic plane and the low metallicity suggest
  that V371 Per may belong to the thick disc (or to the halo)
  of the Galaxy, which is consistent with its low metallicity [Fe/H]=--0.42
  and the enhancement of the $\alpha$- and s-elements relative to iron.
  Line splitting is observed in one of the spectra, which can be due to the
  non-radial pulsations.\\[1mm]
  {\bf Key words}:  Classical Cepheids -- stars: individual: V371 Per \\[2mm]

   {\bf 1. Introduction}\\[1mm]

  Beat Cepheids are classical Cepheid variable stars that simultaneously pulsate in two
  radial modes.
  They are sometimes referred to as double-mode Cepheids.
  A beat Cepheid pulsates either in the first overtone and the fundamental
  modes (P1/P0), or in the second and the first overtone modes
  (P2/P1). Previous studies clearly established that the
  period ratio (higher to lower mode) of the P1/P0 pulsators is
  around 0.72, while that of P2/P0 is closer to 0.80.
  The period ratios can be measured very accurately and have been found to
  correlate with the Cepheid masses, luminosities, \Teff, and the abundances of the heavy elements.
  For example, from the OGLE photometry and the stellar atmosphere models, Kov\'acs
  (2009) showed that in both Magellanic Clouds the average metallicity of
  the P1 Cepheids is lower than those pulsating in the fundamental
  mode.

  Extensive photometry of V371 Per (=BD+41 563 = 2MASS J02553118+4235197)
  over a number of years has clearly shown it to be a Galactic beat Cepheid,
  with the shortest period known so far (P0=1.738 d).
  The high value of the period ratio (P1/P0 = 0.731) suggests
  low metallicity: [Fe/H] should be between --1.0 and --0.7
  according to Wils et al. (2010).
  Its distance, which is derived from
  the empirical period-luminosity (PL) relation, places it in the Galactic thick disk
  or the halo, 0.8 kpc above the Galactic plane.
  The amplitude of the first overtone mode is larger than
  that of the fundamental mode, which is quite rare for the
  Galactic beat Cepheids (Wils et al. 2010).
  Only in AX Vel (with a fundamental
  period of 3.67 d) and V458 Sct (4.84 d) the first
  overtone has a larger amplitude than the fundamental mode.

  In the present paper we report on the detection of the
  Li I 6707.8 \AA \, line in V371 Per.\\[2mm]

  {\bf 2. The spectral material}\\[1mm]

  Four spectra were obtained on three nights in September 2011 with the fiber echelle-type
  spectrograph HERMES, mounted on the 1.2\,m Belgian
  telescope on La Palma. A high-resolution configuration with
  R= 85\,000 and the wavelength coverage 3800--9000 \AA\ was used.
  The spectra were reduced using the Python-based pipe-line,
  that performs the order extraction, wavelength
  calibration using the Thr-Ne-Ar arcs, division by the flat
  field, cosmic-ray clipping, and the order merging.
  For more details on the spectrograph and the pipe-line,
  see Raskin et al (2011).

  We chose to derive abundances from two spectra observed on the same night
  of September 29, because of their superior signal to noise ratio (S/N).
  The rest two spectra were used for the determination of the radial velocity and
  the effective temperature (\Teff, see Table 1).

  We used the DECH 20 software package (Galazutdinov 1992) to normalize the
  individual spectra to the local continuum, to identify the lines of different
  chemical elements, and to measure the equivalent widths of the individual
  lines.\\[2mm]

   \begin{table*}
   \caption[]{Observations of V371 Per, radial velocity measurements
              and photospheric parameters determined in this work.}
   \label{T1}
   \begin{tabular}{ccrcrrrcccc}
   \hline\noalign{\smallskip}
   Spectrum &   HJD     &  $RV$   & $\sigma$&\Teff &$\sigma$& N &\logg &\Vt&[Fe/H]&Remark\\
            & 2455800+  &  (\kms) & (\kms)  &  (K) &  (K)   &   &(\kms)&   &      &      \\
   \noalign{\smallskip}\hline\noalign{\smallskip}
   374513 & 31.6483530&--17.094& 0.112 & 6213 & 320&   8 &... &  ... &  ...&  \\
   374659 & 33.6369571& --2.649& 0.073 & 5984 & 378&  15 &... &  ... &  ...&  \\
   374737 & 34.6177909&--14.455& 0.040 & 5950 & 148&  48 &2.20&  3.70&--.42&+ \\
   374738 & 34.6461523&--12.859& 0.042 & 5996 & 145&  40 &2.20&  3.70&--.42&+ \\
   \noalign{\smallskip}\hline
   \end{tabular}
   \\
   Remark: +:  spectra used for the abundance analysis.
   \end{table*}

   \begin{figure}
   \resizebox{\hsize}{!}{\includegraphics{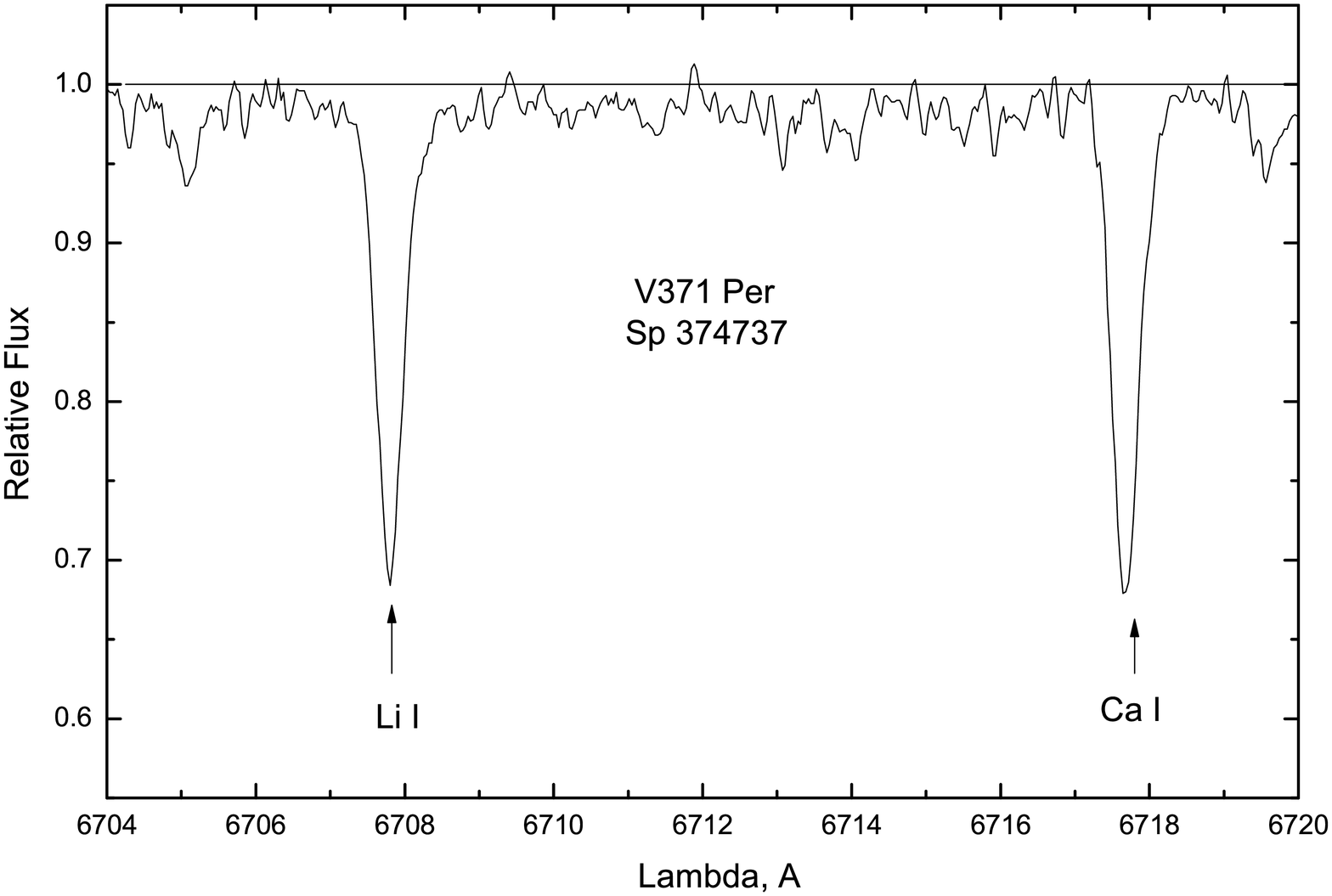}}
   \caption{The Li region in V371 Per spectrum number 374737.}
   \label{Fig1}
   \end{figure}

   \newpage

   {\bf 3. Fundamental parameters and the chemical composition}\\[1mm]

   To determine the effective temperature for our star we employed
   the method of Kovtyukh (2007) which is based on the line depth ratios.
   This technique allows the determination of \Teff\ with an exceptional
   precision. It relies on the ratio of the central depths of two lines that
   have very different functional dependences on \Teff\ (and there are several tens of
   line pairs that are used in this analysis). The method is independent of the
   interstellar reddening and only marginally dependent on individual
   characteristics of stars, such as rotation, microturbulence, metallicity
   and others. The use of $\sim$50 calibrations per spectrum results in
   the uncertainty of 10--20\,K for spectrum with S/N greater than 100,
   and 30--50\,K for S/N less than 100.

   To determine the microturbulent velocities (\Vt) and gravities (\logg), we
   used a modified version of the standard analysis as proposed by Kovtyukh
   \& Andrievsky (1999). In this method the microturbulence is determined from
   {Fe}{\sc ii} lines (instead of {Fe}{\sc i} lines used in the classic abundance analysis).
   The gravity is determined by forcing the equality of the total iron abundance
   derived from {Fe}{\sc i} and {Fe}{\sc ii}. Normally, this method results in
   the iron abundance determined from {Fe}{\sc i} to show a strong
   dependence on the equivalent width (due to the non-LTE effects).
   In this case we take as the proper iron abundance the abundance extrapolated to the zero equivalent width.

   The resulting \Teff, \logg\ and \Vt\ are presented in Table 1.

   The elemental abundances were calculated with the help of the Kurucz's WIDTH9
   code. The resulting averaged values are listed in Table 2. As usual, they are given
   relative to the solar abundances, which were adopted from Grevesse et al. (1996).

   Our oscillator strengths have been obtained by means of the inverse spectroscopic analysis
   of the solar spectrum, namely, by requiring the adopted solar abundance
   for each line with the measured equivalent width (EW).
   The benefit of these "solar" oscillator strengths is that the relative abundances (CNO, in particular) deduced for a given object
   will not change if the the currently adopted solar abundances were to be modified.\\[2mm]

   \begin{figure}
   \resizebox{\hsize}{!}{\includegraphics{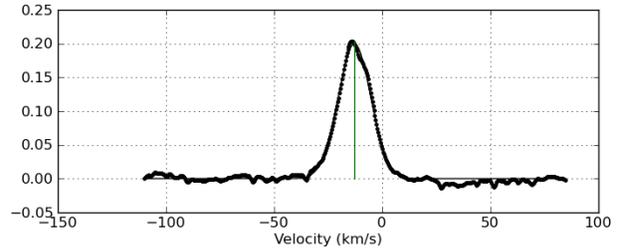}}
   \caption{Cross-correlation function of the spectrum 374738
   with a G2 template. One can see that the average line profile
   in this spectrum consists of at least two absorption components,
   which could be due to the non-radial pulsations.}
   \label{Fig2}
   \end{figure}

   \begin{figure}
   \resizebox{\hsize}{!}{\includegraphics{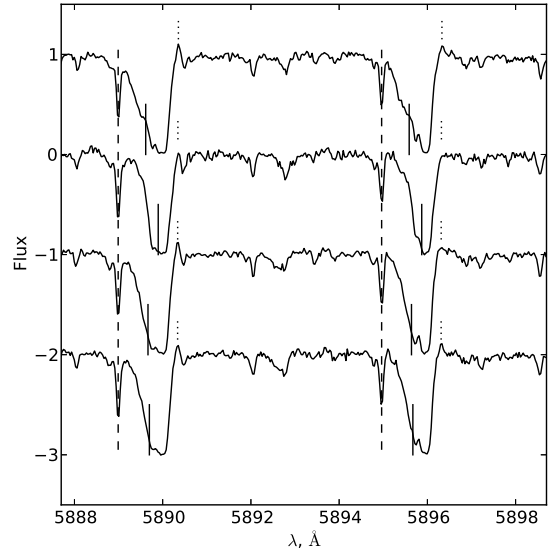}}
   \caption{Complicated profiles of the sodium D12 lines in our spectra of V371 Per.
    Solid lines: photospheric component; dashed: likely interstellar stationary component; dotted: telluric emission.}
   \label{Fig3}
   \end{figure}

   \begin{table}
   \caption[]{Elemental abundances in V371 Per}
   \label{T2}
   \begin{tabular}{lrcrc}
   \hline
    Ion        & [El/H] & $\sigma$ & NL & (El/H) \\
   \noalign{\smallskip}
   \hline\noalign{\smallskip}
   Li {\sc i}  &  2.19 &  ... &  1 & 3.35  \\
   C {\sc i}   &--0.32 & 0.11 &  9 & 8.23  \\
   N {\sc i}   &--0.21 & 0.04 &  2 & 7.76  \\
   O {\sc i}   &--0.18 &  ... &  1 & 8.69  \\
   Na {\sc i}  &--0.45 & 0.17 &  3 & 5.88  \\
   Mg {\sc i}  &--0.43 & 0.00 &  2 & 7.15  \\
   Al {\sc i}  &--0.28 & 0.15 &  4 & 6.19  \\
   Si {\sc i}  &--0.28 & 0.05 & 15 & 7.27  \\
   Si {\sc ii} &--0.30 & 0.16 &  2 & 7.25  \\
   S  {\sc i}  &--0.13 & 0.13 &  6 & 7.08  \\
   Ca {\sc i}  &--0.25 & 0.09 & 11 & 6.11  \\
   Sc {\sc ii} &--0.17 & 0.07 &  9 & 3.00  \\
   Ti {\sc i}  &--0.04 & 0.12 & 32 & 4.98  \\
   Ti {\sc ii} &--0.16 & 0.08 &  9 & 4.86  \\
   V {\sc i}   &--0.24 & 0.09 &  3 & 3.76  \\
   V {\sc ii}  &--0.14 & 0.10 &  4 & 3.86  \\
   Cr {\sc i}  &--0.46 & 0.13 & 14 & 5.21  \\
   Cr {\sc ii} &--0.38 & 0.12 & 13 & 5.29  \\
   Mn {\sc i}  &--0.41 & 0.07 &  3 & 4.98  \\
   Fe {\sc i}  &--0.42 & 0.10 &235 & 7.08  \\
   Fe {\sc ii} &--0.43 & 0.12 & 43 & 7.07  \\
   Co {\sc i}  &--0.27 & 0.14 &  4 & 4.65  \\
   Ni {\sc i}  &--0.34 & 0.09 & 56 & 5.91  \\
   Cu {\sc i}  &--0.35 & 0.17 &  5 & 3.86  \\
   Zn {\sc i}  &--0.08 &  ... &  1 & 4.52  \\
   Y {\sc ii}  &  0.03 & 0.11 &  8 & 2.27  \\
   Zr {\sc ii} &--0.05 & 0.06 &  3 & 2.55  \\
   La {\sc ii} &--0.02 & 0.38 &  2 & 1.20  \\
   Ce {\sc ii} &  0.03 & 0.06 &  6 & 1.58  \\
   Pr {\sc ii} &--0.24 &  ... &  1 & 0.47  \\
   Nd {\sc ii} &--0.22 & 0.10 &  6 & 1.28  \\
   Eu {\sc ii} &  0.01 & 0.05 &  2 & 0.52  \\
   \noalign{\smallskip}\hline
   \end{tabular}
   \begin{list}{}{}
   \item[NL] -- number of lines
   \end{list}
   \end{table}

   {\bf 4. The lithium abundance in V371 Per}\\[1mm]

   For a long time no classical Cepheids or supergiants were known to show the
   Li I 6707.8 \AA \, line. Luck (1982) was the first to identify two lithium supergiants
   in the Galaxy -- HD 172365 and HD 174104. Later on, Luck \& Lambert (1992, 2011) discovered lithium
   in the LMC Cepheid HV 5497 and in Galactic Cepheid V1033 Cyg.
   Every lithium supergiant thus presents a great interest,
   as it may indicate a recent creation or a unique evolutionary path of the object.
   In this paper we present the first detection of the lithium line in all four spectra of
   V371 Per. In Fig. 1 we show the Li region in the spectrum number 374737.

   According to the theory (see de Laverny et al. 2003),
   when a star of about 3 M$_{\odot}$ reaches \Teff = 6400 K,
   lithium starts to be depleted in the photosphere, dropping
   to $\log$N(Li) = 1.0 at about \Teff = 5500 K
   (assuming the original abundance to be equal to the Solar system meteoritic abundance of
   3.3 [Lodders 2003]). This agrees with our estimate of the
   upper limit on $\log$N(Li) = 1.0 for the great majority of Cepheids,
   which follows from the non-detection of the lithium line.
   In contrast, for V371 Per we derive a large overabundance of lithium:
   $\log$N(Li) = 3.35 $\pm$0.09.

   Beside lithium enrichment, V371 Per shows non-symmetrical line profiles
   due to the presence of the additional absorption component (Fig. 2).
   Kovtyukh et al. (2003) proposed that the observed bumps in the line
   profiles in some Cepheid spectra could result from a combination
   of the large broadening (either due to rotation or macroturbulence)
   and the resonant interaction between the radial modes responsible for the
   non-radial oscillations.

   Sodium lines show complicated profiles (Fig. 3):
   photospheric component is overlaid on the saturated,
   likely circumstellar absorption; in addition, there is a narrow
   stationary component at the velocity $\sim$--49 km/s, which can be of
   the interstellar origin.
   \\[2mm]

   {\bf 4. Discussion and conclusions}\\[1mm]

   In an evolved intermediate-mass star one expects the lithium abundance
   to be severely diluted
   due to the combined effects of the mass-loss on the Main Sequence and the
   subsequent first dredge-up.
   The sensitivity to mass-loss stems
   from the fact that in B stars
   (the progenitors of Cepheids) Li remains in only the outer 2\% of the
   star at the end of the Main Sequence.
   Even without the mass loss, the standard stellar evolution predicts a dilution
   about a factor of 60 relative
   to the initial value. Assuming an initial lithium content of
   log A(Li) = 3.3 dex, this means that Cepheids should have
   lithium abundances log A(Li) $<$ 1.5 dex.
   In contrast, V371 Per has a strong
   lithium line with the deduced LTE lithium abundance of log A(Li) = 3.35 dex.
   How could V371 Per maintain such a high abundance of lithium
   in its photosphere?

   The simplest answer is that V371 Per
   is crossing the HR diagram towards the giant branch for the first time.
   The photospheric composition then has not been altered by the dredge-up,
   and we are observing an unaltered abundance of lithium. For this to be the case, the CNO, Na content should
   also be in its original state.
   Indeed, this appears to be true: the [C/Fe], [N/Fe], and [Na/Fe] ratios
   are +0.1, +0.2 and 0.0,
   respectively, while the C/O ratio is 0.72. The [N/Fe] ratio is a bit
   high, but could have been overestimated
   by up to 0.2--0.3 dex due to the non-LTE effects (Lyubimkov et al. 2011).
   The [C/Fe] and C/O
   ratios in V371 Per are significantly larger than the typical ratios of --0.21 and
   0.25, respectively, found in
   Cepheids. They, however,
   are typical of those found in young, unevolved stars.

   Another way to potentially ascertain the evolutionary status of a
   Cepheid is to look for the systematic period change over the time.
   For example, Turner et al. (2010) found four
   first overtone or double mode Cepheids with the period changes:
   Polaris, DX Gem, BY Cas, and HDE 344787.
   They argued that this is a manifestation
   of these stars evolving across the Hertzsprung gap.
   A similar monitoring could help clarify the nature of V371 Per.

   Summarizing, with its peculiar abundances of lithium, carbon, nitrogen and
   sodium (compared with
   ordinary Cepheids) V371 Per can be considered as the Cepheid
   which is presently  crossing the instability strip for the first time.\\[2mm]

  {\it Acknowledgements}

    The spectra were collected with the Mercator Telescope,
   operated on the island
   of La Palma by the Flemish Community, at the Spanish
   Observatorio del Roque de los Muchachos of the Instituto de
   Astrofisica de Canarias.\\[2mm]

   {\bf References\\[2mm]}
   \noindent

   Galazutdinov G.A.: 1992, {\it Preprint SAO RAS}, {\bf 92}, 2

   Grevesse N., Noels A., Sauval J.: 1996, \h {\it ASP Conf. Ser.}, {\bf 99}, 117

   Kov\'acs G.: 2009, {\it EAS Publ. Ser.}, {\bf 38}, 91

   Kovtyukh V.V.: 2007, {\it MNRAS}, {\bf 378}, 617

   Kovtyukh V.V. \& Andrievsky S.M.: 1999, {\it A\&A}, \h {\bf 351}, 597

   Kovtyukh V.V., Andrievsky S.M., Luck R.E., \h Gorlova N.I.:
      2003, {\it A\&A}, {\bf 401}, 661

   de Laverny P., do Nascimento J. D. Jr., Lebre A., \h De Medeiros J. R.:
       2003 {\it A\&A}, {\bf 410}, 937

   Lodders K.: 2003, {\it ApJ}, {\bf 591}, 1220

   Luck R.E.: 1982, {\it PASP}, {\bf 94}, 811

   Luck R.E. \& Lambert D.L.: 1992, {\it ApJSS}, {\bf 79}, 303

   Lyubimkov, L.S., Lambert, D.L., Korotin, S.A., \h Poklad, D.B., Rachkovskaya, T.M., \h
      \& Rostopchin, S.I.: 2011, {\it MNRAS}, {\bf 410}, 1774.

   Raskin G., van Winckel H., Hensberge H. et al.: \h 2011, {\it A\&A}, {\bf 526}, 69

   Turner D. G., Majaess D. J., Lane D. J., \h Percy J. R., English
     D. A., \& Huziak R.: 2010, \h {\it Odessa Astr. Publ.},
     {\bf 23}, 125

   Wils P., Henden A. A., Kleidis S., Schmidt E. G., \h Welch D. L.:
         2010, {\it MNRAS}, {\bf 402}, 1156

  \end{document}